\documentclass[aps,superscriptaddress,showpacs,nofootinbib,eqsecnum,prd]{revtex4} 
                   
\usepackage{epsf}  
\usepackage{eucal} 
\usepackage{amssymb}
\usepackage{amsmath} 
\usepackage{graphicx} 
\usepackage{bm}
\usepackage{dcolumn} 
\usepackage{epsfig} 
\usepackage{multirow}
\usepackage{hyperref}

\begin{document}

\title{Optimized perturbation theory for charged scalar fields at finite
  temperature  and in an external magnetic field}

\author{D. C. Duarte} \email{dyana.c.duarte@gmail.com}
\affiliation{Departamento de Ci\^{e}ncias Naturais, Universidade Federal de
  S\~ao Jo\~ao Del Rei, 36301-000, S\~ao Jo\~ao Del Rei, MG, Brazil}

\author{ R. L. S. Farias} \email{ricardofarias@ufsj.edu.br}
\affiliation{Departamento de Ci\^{e}ncias Naturais, Universidade Federal de
  S\~ao Jo\~ao Del Rei, 36301-000, S\~ao Jo\~ao Del Rei, MG, Brazil}

\author{Rudnei O. Ramos} \email{rudnei@uerj.br} \affiliation{Departamento de
  F\'{\i}sica Te\'orica, Universidade do Estado do Rio de Janeiro, 20550-013
  Rio de Janeiro, RJ, Brazil}

\begin{abstract}

Symmetry restoration in a theory of a self-interacting charged scalar field
at finite temperature and in the presence of an external magnetic field is
examined.  The effective potential is evaluated nonperturbatively in the
context of the optimized  perturbation theory method.  It is explicitly shown
that in all ranges of the  magnetic field, from weak to large fields, the
phase transition is second order  and that the critical temperature increases
with the magnetic field.  In addition, we present an efficient way to deal
with the sum over the Landau levels,  which is of  interest especially in the
case of working with weak magnetic fields.

\end{abstract}


\maketitle

\section{Introduction}

Phase transition phenomena in spontaneously broken quantum field theories have
long been a subject of importance and interest due to their wide range of
possible  applications, going from low energy phenomena in condensed matter
systems to  high energy phase transitions in particle physics and cosmology
(for reviews, see for example~\cite{book,rivers,boya}).

In addition to thermal effects, phase transition phenomena are also known to
be triggered by other external effects, like, for example, by external fields.
In particular, those changes caused by  external magnetic fields have
attracted considerable attention in the past \cite{linde} and received
reinvigorated interest recently, mostly because of the physics associated with
heavy-ion collision experiments. In heavy-ion collisions, it is supposed that
large magnetic fields can be generated, and the study of their effects in the
hadronic phase transition then became subject of intense interest (see
e.g. \cite{recent} for a recent review).  Magnetic fields can lead in
particular to important changes in the chiral/deconfinement transition in
quantum chromodynamics (QCD)~\cite{qcd} and even the possibility of generating
new phases~\cite{efrain}

As far the influence of external magnetic fields and thermal effects on phase
transformations are concerned, one well known example that comes to our mind
is the physics  associated with superconductivity, in particular in the
context of the  Ginzburg-Landau theory~\cite{tink}. Let us recall in that case
thermal effects  alone tend to produce a phase transition at a critical
temperature where  superconductivity is  destroyed and the system goes to a
normal ordered state.  The phase transition in this case is second
order. However, in the presence of an  external magnetic field, but below some
critical value, by increasing the temperature the system undergoes a first
order phase transition instead.  This simple example already shows that
magnetic fields may have influence on the phase transition other than we would
expect from thermal effects alone.  There are also other examples of more
complex systems where external magnetic fields may have a drastic effect on
the symmetry behavior. Among these effects, besides the possibility of
changing the order of the phase transition, as in the Ginzburg-Landau
superconductor,  it can in some circumstances strengthen the order of the
phase transition, like in the electroweak phase transition in the presence of
external fields \cite{Kajantie}, or there can also be dynamical effects, like
delaying the phase transition \cite{Fiore}.  External magnetic fields alone
can also lead to dynamical symmetry breaking (magnetic catalysis)
\cite{Miransky} (for an earlier account, see Ref. \cite{Klimenko}).

Likewise, nonperturbative effects may affect the symmetry properties of a
system, once the external parameters are changed, in a way different than
seeing through a purely perturbative calculation,  or by a mean-field leading
order description.  This is because perturbation theory is typically beset by
problems, for example around critical points, due to infrared divergences, or
at high temperatures, when powers of coupling constants can become surmounted
by  powers of the temperature (see e.g. the textbooks~\cite{bellac,kapusta}
for extensive discussions). Thus, high temperature field theories and the
study of phase transitions in general require the use of nonperturbative
methods, through which large classes of terms need to be resummed . Familiar
techniques used to perform these resummations include, for example,  ring
diagram (or daisy and superdaisy) schemes~\cite{Espinosa,sato}, composite
operator methods~\cite{Camelia} and field propagator dressing
methods~\cite{Banerjee,Parwani}. Other methods used include also numerical
lattice studies and expansions in parameters not related to a coupling
constant, like the $1/N$ expansion and the $\epsilon$-expansion~\cite{zinn},
the use of two-particle irreducible (2PI) effective actions~\cite{cjt,berges},
hard-thermal-loop  resummation~\cite{htl},  variational methods, like
the screened  perturbation theory~\cite{spt,andersen-braaten-strickland} and
the optimized perturbation theory (OPT)~\cite{opt}.  Of course, any
resummation technique must be implemented with care so to avoid  possible
overcounting of terms and lack of self-consistency. Failure in not following
this basic care can lead to a variety of problems, like predicting nonexistent
phenomena or producing a different order for the phase transition. One
classical example of this was the earlier implementations of daisy and
superdaisy schemes,  that at some point were giving wrong results,
e.g. predicting a first order  transition~\cite{carr} for the $\lambda \phi^4$
theory, an unexpected result since  the model belongs to the universality
class of the Ising model, which is second order.   These methods have also
initially predicted a stronger first order phase  transition in the
electroweak standard model, a result soon proved to be
misleading~\cite{arnold}. These wrong results were all because of the wrong
implementation of the ring-diagram summation at the level of the effective
potential, as clearly explained in the first reference in~\cite{arnold}

In this work we will analyze the phase transition for a self-interacting
complex scalar  field model and determine how an external magnetic field,
combined with thermal effects, affects the transition. All calculations will
be performed in the context of the OPT nonperturbative method.  Our reasons
for revisiting here the phase transition in this model are two-fold.  {}First
because this same model has been studied recently in the context of the
ring-diagram resummation  method~\cite{ayala-ring}, where it was found that
the ring-diagrams render the phase transition first order and that the effect
of magnetic fields was to strengthen the order of the transition and also to
lower the critical temperature for the onset of the  (first order) phase
transition.  So in this work we want to reevaluate these findings in the
context of the OPT method.  We recall, from the discussion of the previous
paragraph, that the ring diagram method requires special attention in its
implementation and that previous works have  already concluded  erroneously
about its effects on the transition. The OPT method has a long history of
successful applications (for a far from complete list of previous works and
applications  see e.g. Refs. \cite{previous1,previous2} and references
therein).  The OPT method automatically resums large classes  of terms in a
self-consistent way so to avoid possible dangerous overcounting of diagrams.
In our implementation of the OPT here, we will see that already at the first
order in the OPT it is equivalent to the daisy and superdaisy schemes. The OPT
then provides a safe comparison with these other nonperturbative schemes. In
particular, to our knowledge, this is the first study of the OPT method when
considering the inclusion of an external magnetic field.  {}Finally, we also
want to properly treat the effects of small magnetic fields (which requires
summing over very large Landau-levels) in an efficient way, particularly
suitable for numerical work. This way we can evaluate in a precise way the
effects of external magnetic fields ranging from very small to very large
field intensities  (in which case, in general, just a few Landau-levels
suffice to be considered).  {}For this study we will make use of the
Euler-Maclaurin formula and fully investigate the validity of its use as an
approximation for the Landau level sums for different ranges for the magnetic
field.

The remaining of this paper is organized as follows.  In Sec. II we introduce
the model and explain  the application of the OPT method for the problem we
study in this paper.  It is shown explicitly which  terms are resummed by the
OPT. We also verify that the Goldstone theorem is fulfilled in the OPT.  In
Sec. III we study the phase transition in the spontaneously broken
self-interacting quartic complex scalar field in the OPT method, first by
incorporating only thermal effects and then by including both temperature and
an external magnetic field. In Sec. IV we study the advantages of using  the
Euler-Maclaurin formula for the sum over the Landau levels.  The accuracy and
convergence of the method is fully examined. We determine that  the sum over
the landau levels, in the regime of low magnetic fields, where typically we
must sum over very large levels so to reach good accuracy, can be efficiently
performed within the very few first terms in the Euler-Maclaurin formula. As
an application, an analytical formula for the critical temperature for phase
restoration, as a function of the magnetic field, $T_c(B)$, is derived.  Our
final conclusions are given in Sec. V. {}Finally, an appendix is included to
show some of the details of the renormalization of the model in the context of
the OPT.
    
\section{The complex scalar field model and the OPT implementation}
\label{model+opt}

In our study we will make use of a self-interacting quartic complex scalar
field model with a global $U(1)$ symmetry and  
spontaneously symmetry breaking at tree-level in the
potential, whose Lagrangian density is of the standard form,

\begin{equation}
\mathcal{L}= |\partial_\mu \phi |^2 + m^2 |\phi|^2 - \frac{\lambda}{3 !}
|\phi|^4\;,
\label{lagr}
\end{equation}
where $m^2 >0$ for spontaneously symmetry breaking. It is convenient to write
the complex scalar field $\phi$ in terms of real and imaginary components,
$\phi = (\phi_1 + i\phi_2)/\sqrt{2}$. In terms of a (real) vacuum expectation
value (VEV) for the field, $\langle \phi \rangle \equiv \varphi/\sqrt{2}$, we
can, without loss of generality, shift the field $\phi$ around its VEV in the
$\phi_1$ direction,

\begin{eqnarray}
&&\phi_1 \to \varphi + \phi_1\nonumber \\ 
&&\phi_2 \to \phi_2\;.
\label{shift}
\end{eqnarray}

The {}Feynman propagators for $\phi_1$ and $\phi_2$, in terms of the VEV then
reads,

\begin{eqnarray}
&& D_{\phi_1}(P) =  \frac{i}{P^2 + m^2 - \frac{\lambda}{2}\varphi^2 + i
    \varepsilon}\;,
\label{Dphi1}\\
&& D_{\phi_2}(P) =  \frac{i}{P^2 + m^2 - \frac{\lambda}{6}\varphi^2 + i
  \varepsilon}\;.
\label{Dphi2}
\end{eqnarray}
Using the tree-level VEV for the field, $\varphi_0^2 =  6 m^2/\lambda$ in
Eqs. (\ref{Dphi1}) and (\ref{Dphi2}), $\phi_1$ is then associated with the
massive Higgs mode (with mass squared $m_1^2 =  2 m^2$ at the tree level) and
$\phi_2$ is the Goldstone mode of the field, remaining massless throughout the
symmetry broken phase.
 
\subsection{Implementing the Optimized Perturbation Theory}

The implementation of the OPT  in the Lagrangian density is the standard
one~\cite{previous1,previous2},   where it is implemented through an
interpolation procedure,

\begin{equation}
\mathcal{L} \to \mathcal{L}_{\delta} = \sum_{i=1}^2 \left\{ \frac{1}{2}\left(
\partial_\mu \phi_i\right)^2 - \frac{1}{2}\Omega^{2}\phi_i^2+ \frac{\delta
}{2}\eta^{2}\phi_i^2-\frac{\delta\, \lambda }{4!}\left( \phi_i^2\right) ^{2}
\right\} + \Delta {\cal L}_{\rm ct, \delta}\;,
\label{interpol}
\end{equation}
where $\Omega^2=-m^2 + \eta^2$ and $\Delta {\cal L}_{\rm ct,\delta}$ is the
Lagrangian density part with the renormalization counterterms needed to render
the theory finite.  In $\mathcal{L}_{\delta}$, the dimensionless parameter
$\delta$ is a bookkeeping parameter used only to keep track of the order that
the OPT is implemented (it is set to one at the end)  and $\eta$ is a (mass)
parameter determined variationally at any given finite order of the OPT. A
popular variational criterion used to determine $\eta$ is known as the
principle of minimal sensitivity  (PMS), defined by the variational
relation~\cite{PMS}

\begin{equation}
\frac {d \Phi^{(k)}}{d \eta}\Big |_{\bar \eta, \delta=1} = 0 \;,
\label{pms}
\end{equation} 
which is applied to some physical quantity $\Phi^{(k)}$, calculated up to some
order-$k$ in the OPT. The optimum value $\bar \eta$, which satisfies
Eq.~(\ref{pms}),  is a function of the original parameters of the theory and
it is in general a nontrivial function of the couplings. It is because of the
variational principle used that nonperturbative results are generated. Other
variational criteria can  likewise be defined differently, but they produce
the same final result for  the quantity $\Phi^{(k)}$, as shown
recently~\cite{FGR}.

In terms of the interpolated Lagrangian density, Eq. (\ref{interpol}),  the
{}Feynman rules in the OPT method are as follows. The interaction vertex is
changed from $-i \lambda$ to $-i \delta \lambda$. The quadratic terms  $
\delta \eta^2 \phi_i^2/2$  in Eq. (\ref{interpol}) define new vertex insertion
terms. The propagators for $\phi_i$ ($i=1,2$) now become

\begin{eqnarray}
&& D_{\phi_1,\delta}(P) =  \frac{i}{P^2 - \Omega^2  - \frac{\delta
      \lambda}{2}\varphi^2 + i \varepsilon}\;,
\label{Dphi1-delta}\\
&& D_{\phi_2,\delta}(P) =  \frac{i}{P^2 - \Omega^2  - \frac{\delta
    \lambda}{6}\varphi^2 + i \varepsilon}\;.
\label{Dphi2-delta}
\end{eqnarray}

\subsection{The effective potential in the OPT method}
\label{OPT+Veff}

\begin{figure}[htb]
  \vspace{0.75cm} \epsfig{figure=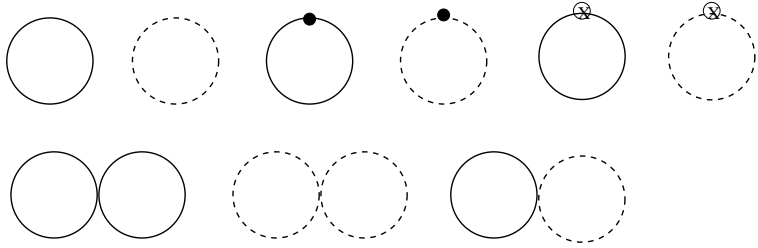,angle=0,width=10cm}
\caption[]{Feynman diagrams contributing to the effective potential to first
  order in the OPT. Solid lines stand for the $\phi_1$ propagator, while
  dashed lines stand for the $\phi_2$ propagator. A black dot is an insertion
  of $\delta \eta^2$. A crossed dot is a mass renormalization insertion.}
\label{vacuum}
\end{figure}

As in the many other previous applications~\cite{previous2,FGR}, we apply the
PMS,  Eq. (\ref{pms}), directly on the effective potential, which is the most
convenient quantity to study the phase structure of the model. To first order
in the OPT, using the  previous {}Feynman rules in the OPT method given above,
the effective potential $V_{\rm eff}(\varphi)$ is given by the vacuum diagrams
shown in {}Fig. \ref{vacuum}.

{}From the diagrams shown in {}Fig. \ref{vacuum} and by further expanding the
propagators Eqs. (\ref{Dphi1-delta}) and (\ref{Dphi2-delta}) in $\delta$, the
explicit expression for the renormalized  effective potential at first order
in the OPT  (some of the details of the renormalization in the OPT method are
given in the  Appendix A) becomes 

\begin{eqnarray}
V_{\rm eff} (\varphi) &=& \frac{\Omega ^{2}}{2}\varphi^{2} - \delta
\frac{\eta^2}{2}  \varphi^2 + \delta\frac{ \lambda }{4!}\varphi^{4} \nonumber
\\ &-& i\int_P \ln(P^2 - \Omega^2 ) - \delta \eta^2 \int_P \frac{i}{P^2 -
  \Omega^2} + \delta  \frac{\lambda}{3}\varphi^2 \int_P \frac{i}{P^2 -
  \Omega^2} \nonumber \\ &+& \delta  \frac{\lambda}{24 \pi^2}
\frac{1}{\epsilon} \Omega^2 \int_P \frac{i}{P^2 - \Omega^2} + \delta
\frac{\lambda}{3} \left[\int_P \frac{i}{P^2 - \Omega^2}\right]^2\;,
\label{Veff}
\end{eqnarray}
where the momentum integrals are expressed in Euclidean space and the
regularization is performed in $\overline{\mathrm{MS}}$ scheme, where, in the
finite temperature only case,

\begin{equation}
\int_P \equiv iT \sum_{P_0=i\omega_n} \left(\frac{e^{\gamma_E}
  M^2}{4\pi}\right)^\epsilon \int\frac{d^{d}p}{(2\pi)^d}\;,
\label{intT}
\end{equation}
where $\gamma_E$ is the Euler-Mascheroni constant, $M$ is an arbitrary mass
regularization scale, $d=3-2\epsilon$ is the dimension of space and $\omega_n
= 2 \pi n T, \, (n=0, \pm 1, \cdots)$ are the Matsubara frequencies for bosons
at temperature $T$.

The inclusion of an external magnetic field presents no additional difficulty.
{}For example, without loss of generality, we can consider a constant magnetic
field in the $z$-space direction and use a gauge where the external
electromagnetic field is $A_\mu = (0,0,B x,0)$.  The {}Feynman propagator for
a charged scalar field with charge $e$ becomes $P^2 - m^2 \to  -\omega_n^2 -
E_k^2(p_z,B)$, where $E_k(p_z,B)$ is the energy dispersion (for charged scalar
bosons) in an external constant magnetic field~\cite{tsai},

\begin{equation}
E_k^2 = p_z^2 + m^2 + (2 k+1) eB,
\label{Ek}
\end{equation}
where the last term denotes the Landau levels ($k=0, 1, 2, \cdots$).
Likewise, the momentum integrals, taking into account the degeneracy
multiplicity of the Landau levels~\cite{schwinger}, are now represented by

\begin{equation}
\int_P \equiv i\frac{e B}{2 \pi} \sum_{k=0}^{+\infty} T \sum_{P_0=i\omega_n}
\left(\frac{e^{\gamma_E} M^2}{4\pi}\right)^\epsilon
\int\frac{d^{d-2}p_z}{(2\pi)^{d-2}}\;.
\label{intTB}
\end{equation}

\subsection{Optimization results and Goldstone theorem}

Let us now apply the PMS condition (\ref{pms}) on the effective potential
Eq. (\ref{Veff}).  We obtain straightforwardly that the optimum $\bar \eta$
satisfies the nontrivial (renormalized) gap equation

\begin{eqnarray}
{\bar \eta}^2   &=& \frac{\lambda}{3} \varphi^2 + \frac{2\lambda}{3}
\left[\int_P \frac{i}{P^2 - \Omega^2} + \frac{\Omega^2}{16 \pi^2}
  \frac{1}{\epsilon} \right]\Bigr|_{\eta = {\bar \eta}}\;.
\label{etabar}
\end{eqnarray}

The extrema of the effective potential are defined as usual, by requiring that
the first derivative of the effective potential with respect to $\varphi$
vanishes. This gives us the trivial solution ${\bar \varphi}=0$ and the
solution for the minimum,

\begin{equation}
{\bar \varphi}^2 = 6\frac{m^2}{\lambda} -4 \left[ \int_P \frac{i}{P^2 -
    \Omega^2} + \frac{\Omega^2}{16 \pi^2} \frac{1}{\epsilon} \right]\;.
\label{varphi}
\end{equation}

We can now verify the effective mass for the field, in particular for the
$\phi_2$ component of the complex scalar field. The effective mass for
$\phi_2$ in the OPT is given by

\begin{equation}
m_{\rm eff,2}^2 = -m^2 + {\bar \eta}^2 + \frac{\lambda}{6}{\bar \varphi}^2 +
\Sigma_2({\bar \eta})\;,
\label{m2}
\end{equation}
where $\Sigma_2({\bar \eta})$ is the (renormalized) field's self-energy, which
at first order in the OPT is trivially found to be given by

\begin{equation}
\Sigma_2^{(1)}({\bar \eta}) = -{\bar \eta}^2 + \frac{2\lambda}{3} \left[
  \int_P \frac{i}{P^2 - \Omega^2} + \frac{\Omega^2}{16 \pi^2}
  \frac{1}{\epsilon} \right]\Bigr|_{\eta = {\bar \eta}}\;.
\label{Sigma1}
\end{equation}

Using now Eqs. (\ref{Sigma1}), (\ref{etabar}) and (\ref{varphi}) in
Eq. (\ref{m2}), we obtain immediately that $m_{\rm eff,2}^2=0$, which is
nothing but the result expected due to Goldstone theorem for the symmetry
broken complex scalar field model.  It can also be verified that this result
carries out at higher orders in the OPT method, in which case, at some
order-$k$ in the OPT, the self-energy is the one at the respective order,
$\Sigma^{(k)}$, entering in the above equations.  Previous demonstrations of
the Goldstone theorem in the OPT were for the linear sigma
model~\cite{hatsuda} and for the $SU(2)$ Nambu--Jona-Lasinio model~\cite{njl}.

{}Finally, it is noticed from the PMS equation given above,
Eq. (\ref{etabar}), that the OPT naturally resums the leading order loop terms
of the field's self-energy. So it is quite analogous, at already in the first
order in the OPT approximation,  to the ring-diagram
resummation~\cite{Espinosa,arnold}.  And that this resummation is also
performed automatically in a self-consistent way, it is clear from the
interpolation procedure.  While ${\bar \eta}$ enters in all propagators, thus
carrying the self-energy corrections,  the insertions of ${\bar \eta}$ (the
term $\delta \eta^2 \phi_i^2/2$ in  Eq. (\ref{interpol}) is treated as an
additional interaction vertex) subtract  spurious contributions at each order,
thus avoiding dangerous overcounting of diagrams (this is also similar to the
procedure used in the  screened perturbation
theory~\cite{spt,andersen-braaten-strickland}).  Going to higher orders in the
OPT resums higher order classes of diagrams.  {}For our purposes of comparing
with the ring-diagram results, it suffices then to keep up to first order in
the OPT.
  
\section{Phase structure and symmetry restoration at finite 
temperature and in an external magnetic field}

We are now ready to study the phase structure and the symmetry restoration  in
the complex scalar field model at finite temperature and in an external
magnetic field. It is convenient to first investigate the case of finite
temperature only, so to later compare with the case when an external magnetic
field is added.

\subsection{Symmetry restoration at finite temperature}

It is convenient to write the explicit expressions for each term entering in
the effective potential in Eq. (\ref{Veff}). At finite temperature and in the
absence of  an external magnetic field, using (\ref{intT}), we have that the
first momentum integral in (\ref{Veff}) becomes (recalling that $\Omega^2 =
-m^2 +\eta^2$)

\begin{eqnarray}
-i \int_P \ln(P^2 - \Omega^2) = - \frac{\Omega^4}{2 (4 \pi)^2}
\frac{1}{\epsilon} + Y(T,\eta)\;,
\label{ln}
\end{eqnarray}
where we have identified explicitly the divergent term and the finite term,
$Y(T,\eta)$, is given by

\begin{eqnarray}
Y(T,\eta) &=& - \frac{1}{2 (4 \pi)^2} \left[ \frac{3}{2} +
  \ln\left(\frac{M^2}{\Omega^2}\right) \right] \Omega^4 \nonumber \\ 
&+& \frac{T^4}{\pi^2} \int_0^\infty dz \, z^2
\ln\left[1-\exp\left(-\sqrt{z^2+\Omega^2/T^2}\right)\right]\;.
\label{YT}
\end{eqnarray}
Likewise, for the remaining momentum integrals in (\ref{Veff}) we obtain

\begin{eqnarray}
-\delta \eta^2  \int_P \frac{i}{P^2 - \Omega^2} &=&  -\delta \eta^2
\left[-\frac{\Omega^2}{(4\pi)^2} \frac{1}{\epsilon} + X(T,\eta)\right]\;,
\label{Fb}
\\ 
\delta \frac{\lambda}{3} \varphi^2  \int_P \frac{i}{P^2 - \Omega^2} &=&
\delta \frac{\lambda}{3} \varphi^2   \left[-\frac{\Omega^2}{(4\pi)^2}
  \frac{1}{\epsilon} + X(T,\eta)\right]\;,
\label{Fc}
\end{eqnarray}
where $X(T,\eta)$ is given by

\begin{equation}
X\left( T,\eta\right) =\frac{\Omega^{2}}{16\pi ^{2}}\left[ \ln \left(
  \frac{\Omega^{2}}{ M^{2}}\right) -1\right] +\frac{T^2}{2 \pi^2}\int_0^\infty
dz \frac{z^2}{\sqrt{z^2+\frac{\Omega^2}{T^2}}}\;\frac{1}{
  \exp\left(\sqrt{z^2+\frac{\Omega^2}{T^2}}\right)-1}\;.
\label{XT}
\end{equation}
Next, there is the term coming from the mass counterterm in Eq. (\ref{Veff})
(see Appendix A), 

\begin{eqnarray}
\delta  \frac{\lambda}{24 \pi^2} \frac{1}{\epsilon} \Omega^2 \int_P
\frac{i}{P^2 - \Omega^2} &=& -\frac{2\delta \lambda }{3( 4\pi )^{4}}\,
\Omega^{4} \frac{1}{\epsilon ^{2}} +\frac{\delta \lambda }{24\pi^{2}} \Omega^2
X\left(T,\eta\right) \frac{1}{\epsilon } \nonumber \\ &-&\delta \frac{2
  \lambda}{3}\frac{ \Omega^{4}}{\left( 4\pi\right)^{4}}\,W( \eta) \;,
\label{Fd}
\end{eqnarray}
where we have also defined

\begin{equation}
W(\eta) =\frac{1}{2}\left[ \ln \left( \frac{\Omega^{2}}{ M^{2}}\right)
  -1\right] ^{2}+\frac{1}{2}+\frac{\pi ^{2}}{12}\;.
\label{W0}
\end{equation}

{}Finally, we have the two-loop contributions shown in {}Fig. \ref{vacuum}.
They all sum up in the OPT expansion to give

\begin{eqnarray}
\delta \frac{\lambda}{3} \left[\int_P \frac{i}{P^2 - \Omega^2}\right]^2 &=&
\delta \frac{\lambda}{3} \frac{\Omega^{4}}{\left(
  4\pi\right)^{4}}\frac{1}{\epsilon ^{2}} -\delta \lambda \frac{ \Omega^{2} }{
  24\pi^{2}} X\left( T,\eta\right) \frac{1}{\epsilon } \nonumber \\ &+&\delta
\frac{2\lambda}{3} \frac{\Omega^{4}}{\left( 4\pi\right)^{4}} W(\eta) +  \delta
\frac{\lambda}{3}  X^{2}\left( T,\eta\right)\;.
\label{Fe}
\end{eqnarray}

{}From Eqs. (\ref{ln}), (\ref{Fb}), (\ref{Fc}), (\ref{Fd}) and (\ref{Fe}), we
find the complete expression for the renormalized effective potential at first
order in the OPT:

\begin{eqnarray}
V_{\rm eff}(\varphi,T,\eta) &=& -\frac{m^2}{2}\varphi^{2} + (1-\delta)
\frac{\eta^2}{2}  \varphi^2 + \delta\frac{ \lambda }{4!}\varphi^{4} 
\nonumber \\ 
&+& Y(T,\eta) + \delta \left\{- \eta^2 + \frac{\lambda}{3} \left[\varphi^2
  +  X(T,\eta)\right]\right\} X\left( T,\eta\right)\;.
\label{VeffT}
\end{eqnarray}

The phase structure at finite temperature is completely determined by
Eqs. (\ref{VeffT}), (\ref{etabar}) and (\ref{varphi}). The optimum ${\bar
  \eta}$, determined by the PMS criterion, and the vacuum expectation value
(VEV) for the scalar field,  can now be both, respectively, be expressed as

\begin{eqnarray}
\bar{\eta}^2 = \frac{\lambda}{3} \varphi^2 + \frac{2\lambda}{3} 
X(T,{\bar \eta})\;,
\label{etabarT}
\end{eqnarray}
and

\begin{equation}
{\bar \varphi}^2 = 6\frac{m^2}{\lambda} -4 X(T,{\bar \eta})\;.
\label{varphiT}
\end{equation}

In the OPT we can exactly compute the critical temperature of phase
transition. Using that at the critical point the VEV of the field vanishes,
${\bar \varphi}(T_c) =0$, then from Eq. (\ref{etabarT}) we obtain that

\begin{eqnarray}
\bar{\eta}^2(T_c) = \frac{2\lambda}{3} X(T_c,{\bar \eta}(T_c))\;,
\label{etabarTc}
\end{eqnarray}
which upon using it in Eq. (\ref{varphiT}), we obtain

\begin{equation}
0 = 6\frac{m^2}{\lambda} -4 X(T_c,{\bar \eta}(T_c)) \Rightarrow m^2 -
\bar{\eta}^2(T_c) = - \Omega^2(T_c) =0 \;.
\label{varphiTc}
\end{equation}
{}From the definition of  $X(T,{\bar \eta})$, Eq. (\ref{XT}), we  then obtain
that

\begin{equation}
X(T_c,{\bar \eta}(T_c)) = \frac{T_c^2}{2 \pi^2}\int_0^\infty dz \frac{z}{
  \exp\left(z\right)-1} = \frac{T_c^2}{12}\;,
\label{XTc}
\end{equation}
and when using the above result back in Eq. (\ref{varphiTc}), we obtain the
exact result for the critical temperature at this first order in the OPT:

\begin{equation}
T_c^2 = \frac{18 m^2}{\lambda}\;.
\label{Tc}
\end{equation}
This result for $T_c$ is the same predicted before for a scalar field model
with two real components~\cite{jackiw}, obtained in the high temperature 
one-loop approximation. The result (\ref{Tc}) is exact in our OPT
approximation and obtained independent of any high temperature approximation,
as usually  assumed in any previous calculations. 


\begin{figure}[htb]
  \vspace{0.75cm} \epsfig{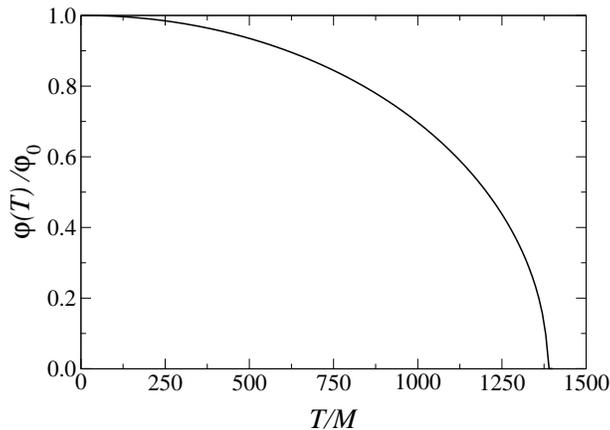}
\caption{Temperature dependence of the minimum of the effective potential
  $\varphi(T)$ in the OPT with the PMS optimization criterion. $\varphi_0$ is
  the  minimum of the tree-level potential. The parameters used are $m/M=20$
  and $\lambda=0.00375$.}
\label{fig2}
\end{figure}

\begin{figure}[htb]
  \vspace{0.75cm} \epsfig{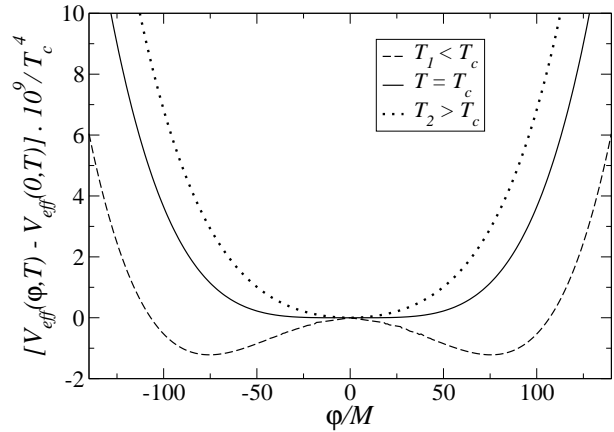}
\caption{The effective potential (subtracting the vacuum
  energy at $\varphi=0$),  
  for the same parameters of {}Fig. \ref{fig2} and for temperatures:
  $T_1/M = 1380$, $T_2/M=1390$ and at the critical temperature, $T_c/M \simeq
  1385.64$.}
\label{fig3}
\end{figure}

{}From Eqs. (\ref{etabarT}) and (\ref{varphiT}), we obtain directly the
behavior of the VEV ${\bar \varphi}$ as a function of the temperature. {}For
comparison purposes, we use analogous parameters as adopted in
Ref. \cite{ayala-ring} \footnote{ Note that in \cite{ayala-ring} the authors
  used a different numerical factor for the quartic term in the tree-level
  potential, which gives rise to the extra factor $3/2$ in the numerical value
  for $\lambda$ in our notation.  In the approximations used in
  \cite{ayala-ring}, the terms involving the regularization scale were not
  included.  Here we express all quantities in terms of the regularization
  scale $M$.}, where $m/M=20$ and $\lambda = 0.00375$. {}For these parameters,
the authors in  Ref.~\cite{ayala-ring} find a first order phase transition.
In {}Fig. \ref{fig2} we show the result for the minimum of the effective
potential, ${\bar \varphi} \equiv \varphi(T)$   as a function of the
temperature. The effective potential, for different temperatures and same
parameters as in {}Fig. \ref{fig2}, is shown in {}Fig. \ref{fig3}.  Note that
the effective potential has an imaginary part for temperatures below $T_c$,
Eq. (\ref{Tc}), since there can be values of $\varphi$ for which $-m^2 + {\bar
  \eta}^2$ becomes negative. This happens for values of $\varphi$ in between
the inflection points  of the potential, which defines the spinodal region of
instability in between the (degenerate) VEV of the field, determined by
Eq. (\ref{varphiT}). In {}Fig. \ref{fig3}, for the case of $T<T_c$, we show
the real part of the effective potential, so to be able to show all the
potential, including the spinodal region. 


It should be noted that the minimum of the effective potential, determined by
the coupled Eqs. (\ref{etabarT}) and (\ref{varphiT}), is well defined and
unique, where the only solutions for Eq. (\ref{varphiT}) are the two
degenerate minima for $T<T_c$, while for $T\geq T_c$, the only solution is
${\bar \varphi} =0$.  Therefore, the phase transition can readily be inferred
to be second order for any set of parameters.  It is also clear from the
results shown by  {}Figs. \ref{fig2} and \ref{fig3} that the transition due to
thermal effects only is of second order. The effective potential  does not
develop local minima and the VEV of the field varies continuously from  $T=0$
till $T=T_c$, where it vanishes. This is the expected behavior, since the
complex scalar field model belongs to the same universality class of a $O(2)$
Heisenberg model, for which the phase transition is second order~\cite{book}.

Thus, we see that the OPT method correctly describes the phase transition in
the model, capturing the correct physics. {}Furthermore, as an added bonus, we
can exactly compute the critical temperature of transition  

\subsection{The phase structure in a constant external magnetic field}

Let us now investigate the phase structure when an external magnetic field is
also applied to the system. In the presence of a constant magnetic field, the
{}Feynman rules  change as shown in Subsec. \ref{OPT+Veff}. The momentum
integrals in Eq. (\ref{Veff}) are given by (\ref{intTB}) and the dispersion
relation for charged particles is given by Eq. (\ref{Ek}), which takes into
account the Landau energy  levels of a charged particle in a magnetic
field\footnote{Note that the dispersion relation Eq. (\ref{Ek}) only applies
  for the fields in the complex base $(\phi,\phi^*)$, since $(\phi_1,\phi_2)$
  are not  appropriate eigenstates of charge. However, when re-expressing the
  effective potential in the complex scalar field base, the final result turns
  out to be  the same as in Eq. (\ref{Veff}) or (\ref{VeffT}),  but with the
  functions $X$ and $Y$ depending now also on the magnetic field.}. 

At finite $B$, the first momentum integral term in (\ref{Veff}) becomes:

\begin{eqnarray}
-i \int_P \ln(P^2 - \Omega^2) = \frac{e B}{2 \pi} \sum_{k=0}^{+\infty}
\left(\frac{e^{\gamma_E} M^2}{4\pi}\right)^\epsilon \int\frac{d^{1-2
    \epsilon}p_z}{(2\pi)^{1-2 \epsilon}} T \sum_{n=-\infty}^{+\infty}
\ln\left[ \omega_n^2 + p_z^2 + \Omega^2 + (2 k+1)eB \right]\;.
\label{ln1}
\end{eqnarray}
By performing the sum over the Matsubara frequencies in (\ref{ln1}), we obtain
two terms.  One is independent of the temperature and the other is for $T\neq
0$. The $T=0$ term is

\begin{eqnarray}
&&\frac{e B}{2 \pi} \sum_{k=0}^{+\infty} \left(\frac{e^{\gamma_E}
    M^2}{4\pi}\right)^\epsilon  \int \frac{d^{1-2 \epsilon}p_z}{(2 \pi)^{1-2
      \epsilon}} \sqrt{ p_z^2 + \Omega^2 + (2 k+1)eB} \nonumber \\ && = -
  \frac{(e B)^2}{4 \pi^2} \left(\frac{e^{\gamma_E} M^2}{8\pi
    eB}\right)^\epsilon \frac{\Gamma(-1+\epsilon)}{(4\pi)^{-\epsilon}}
  \zeta\left(-1+\epsilon, \frac{\Omega^2 + eB}{2 eB}\right) \nonumber \\ &&=
  -\frac{\Omega^4}{32 \pi^2}\left[ \frac{1}{\epsilon} +1 +\ln \left(
    \frac{M^2}{2 eB}\right) +{\cal O}(\epsilon) \right]  + \frac{(eB)^2}{4
    \pi^2} \zeta'\left(-1,\frac{\Omega^2 + eB}{2 eB}\right),
\label{ln1T0}
\end{eqnarray}
where to write the second line in (\ref{ln1T0}), we used the analytic
continuation of the Hurwitz zeta function~\cite{zeta},

\begin{equation}
\zeta(s,a) = \sum_{k=0}^{\infty} \frac{1}{(k+a)^s}\;,
\label{zetafunc}
\end{equation}
and $\zeta'(s,a)$ is its $s$-derivative. In writing the final terms in
Eq. (\ref{ln1T0}) we have also dropped constant terms since they do not affect
the phase structure. 

The $T\neq 0$ part of (\ref{ln1}) is

\begin{eqnarray}
\frac{e B}{\pi} T^2 \sum_{k=0}^{+\infty} \int_{-\infty}^{+\infty} \frac{d z}{2
  \pi} \ln \left\{ 1- \exp\left[ - \sqrt{z^2 + \frac{\Omega^2}{T^2} +  (2k +
    1)\frac{e B}{T^2}} \right] \right\}\;.
\label{ln1T}
\end{eqnarray}
We can now write Eq. (\ref{ln1}) in a form analogous to Eq. (\ref{ln}),

\begin{eqnarray}
-i \int_P \ln(P^2 - \Omega^2) = - \frac{\Omega^4}{2 (4 \pi)^2}
\frac{1}{\epsilon} + {\tilde Y}(T,B,\eta)\;,
\label{lnTB}
\end{eqnarray}
where ${\tilde Y}(T,B,\eta)$ is given by

\begin{eqnarray}
{\tilde Y}(T,B,\eta) &=& -\frac{\Omega^4}{32 \pi^2}\left[1 +\ln \left(
  \frac{M^2}{2 eB}\right) \right] + \frac{(eB)^2}{4 \pi^2}
\zeta'\left(-1,\frac{\Omega^2 + eB}{2 eB}\right) \nonumber \\ &+& \frac{e
  B}{\pi} T^2 \sum_{k=0}^{+\infty} \int_{-\infty}^{+\infty} \frac{d z}{2 \pi}
\ln \left\{ 1- \exp\left[ - \sqrt{z^2 + \frac{\Omega^2}{T^2} +  (2k +
    1)\frac{e B}{T^2}} \right] \right\}\;.
\label{YTB}
\end{eqnarray}

Analogous manipulations leading to Eq.~(\ref{lnTB}) allow us to write the
remaining momentum integral terms in Eq.~(\ref{Veff}) like

\begin{eqnarray}
\int_P \frac{i}{P^2 - \Omega^2} =  -\frac{\Omega^2}{(4\pi)^2}
\frac{1}{\epsilon} + {\tilde X}(T,B,\eta)\;,
\label{tadloop}
\end{eqnarray}
where

\begin{eqnarray}
{\tilde{X}}(T,B,\eta) &=& \frac{e B}{8 \pi^2} \ln \left[ \Gamma \left( \frac{
    \Omega^2 + e B}{2 e B} \right) \right] -\frac{e B}{16 \pi^2} \ln(2 \pi) -
\frac{\Omega^2}{16 \pi^2} \ln\left( \frac{M^2}{2 e B} \right)\nonumber \\ &+&
\frac{e B}{2 \pi} \sum_{k=0}^{+\infty} \int_{-\infty}^{+\infty} \frac{d z}{2
  \pi} \frac{1}{\sqrt{z^2 + \frac{\Omega^2}{T^2} +  (2k + 1)\frac{e
      B}{T^2}}}\; \frac{1}{\exp\left[\sqrt{z^2 + \frac{\Omega^2}{T^2} +  (2k +
      1)\frac{e B}{T^2}}\right] -1}\;.
\label{XTB}
\end{eqnarray}
Actually, Eq. (\ref{tadloop}) is just the derivative of Eq. (\ref{lnTB}) with
respect to $\Omega^2$, as can be easily checked.

The final expression for the renormalized  effective potential at finite
temperature and  in a constant external magnetic field has an analogous form
as Eq. (\ref{VeffT}), where by  using Eqs. (\ref{lnTB}) and (\ref{tadloop}) in
(\ref{Veff}), we obtain

\begin{eqnarray}
V_{\rm eff}(\varphi,T,B,\eta) &=& -\frac{m^2}{2}\varphi^{2} + (1-\delta)
\frac{\eta^2}{2}  \varphi^2 + \delta\frac{ \lambda }{4!}\varphi^{4} \nonumber
\\ &+& {\tilde Y}(T,B,\eta) + \delta \left\{- \eta^2 + \frac{\lambda}{3}
\left[\varphi^2 +  {\tilde{X}}(T,B,\eta)\right]\right\}
     {\tilde{X}}(T,B,\eta)\;.
\label{VeffTB}
\end{eqnarray}

{}Finally, the PMS criterion and the minimum of the effective potential are
again also of the form as Eqs. (\ref{etabarT}) and~(\ref{varphiT}):

\begin{eqnarray}
\bar{\eta}^2 = \frac{\lambda}{3} \varphi^2 + \frac{2\lambda}{3}
    {\tilde{X}}(T,B,{\bar \eta})\;,
\label{etabarTB}
\end{eqnarray}
and

\begin{equation}
{\bar \varphi}^2 = 6\frac{m^2}{\lambda} -4 {\tilde{X}}(T,B,{\bar \eta})\;.
\label{varphiTB}
\end{equation}

\begin{figure}[htb]
  \vspace{0.75cm} \epsfig{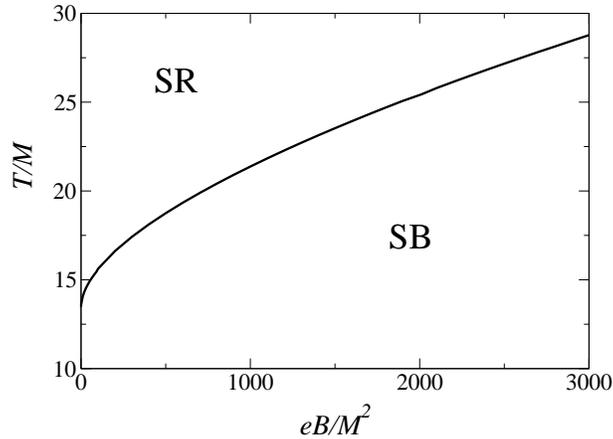}
\caption[]{The phase diagram of the system in the $(B,T)$ plane.  The
  parameters considered here are: $\lambda = 0.1$, $m/M = 1$. The solid line
  separates the regions of symmetry broken (SB) and symmetry restored (SR)
  phases.}
\label{fig4}
\end{figure}

With Eqs. (\ref{VeffTB}), (\ref{etabarTB}) and (\ref{varphiTB}), we are now in
position to study the effects of an external magnetic field in the phase
structure of the model, in addition to the thermal effects.  In
{}Fig. \ref{fig4} we show the phase diagram for the symmetry broken  complex
scalar field model in the $(B,T)$ plane. It shows that the magnetic field
strengths the symmetry broken phase, increasing the critical temperature for
phase transition. 
We must note that this result for the phase diagram is very different 
from that seen in the
superconductor or in the scalar quantum electrodynamics (QED) case
\cite{scalarQED}. In contrast to our case, where we are only studying 
the effects of $B$ and $T$ on the charged scalars, in the scalar QED there is also
the interactions with the gauge field.
In the scalar QED case with a local $U(1)$ gauge symmetry, in the symmetry broken phase,
because of the Higgs mechanism the gauge field
becomes massive and the external magnetic field gets screened (the Meissner
effect). As the magnetic field is increased, eventually above a critical
field the phase is restored through a first order phase transition.
In the study made in this paper, on the other hand, we consider only the
case of a broken {\it global} $U(1)$ symmetry\footnote{One of the authors (ROR) would
like to thank K. G. Klimenko for discussions on this topic and for also
pointing him out the
possible similarity with the symmetry behavior found in this paper due to the magnetic field,
with pion condensation seen in the Nambu--Jona-Lasinio  model with 2 flavor quarks plus baryon 
and isospin chemical potentials~\cite{pioncond}.}.

\begin{figure}[htb]
  \vspace{0.75cm} \epsfig{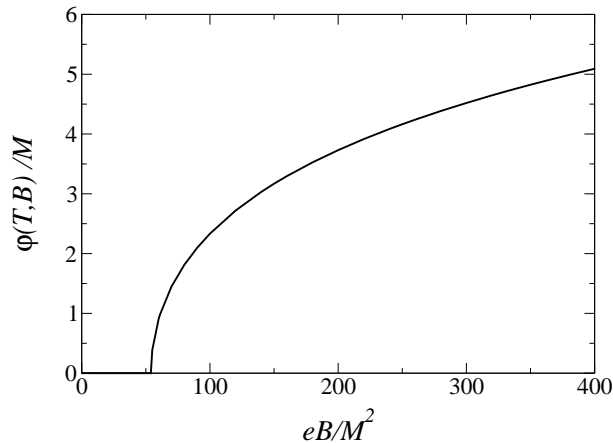}
\caption[]{Magnetic field dependence of the VEV of the field, ${\bar
    \varphi}(T,B)$, for a fixed value of $T$ above $T_c$.  The parameters
  considered here are: $\lambda = 0.1$, $m/M = 1$ and $T/M = 15$.}
\label{fig5}
\end{figure}

\begin{figure}[!htb]
  \vspace{0.75cm} \epsfig{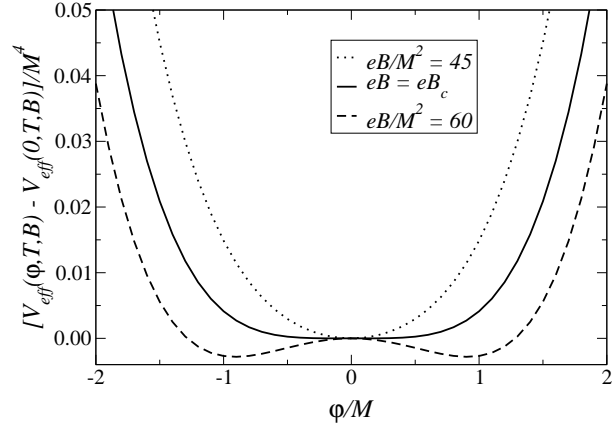}
\caption[]{The effective potential for the same parameters of Fig. \ref{fig5}
  and for three different values of the magnetic field: $e B/M^2 = 45$, $e
  B_c/M^2 = 53.93$  and  $e B/M^2=60$.}
\label{fig6}
\end{figure}

The effect of the magnetic field on the global $U(1)$ symmetry can 
also be seen in the next two figures. In
{}Fig. \ref{fig5} we have plotted the VEV of the field at fixed $T>T_c$, thus
starting from a symmetry restored phase, ${\bar \varphi} =0$, as a function
of $B$. The critical temperature, for the parameters used, is $T_c/M \simeq
13.42$, which fully agrees with Eq. (\ref{Tc}), and the symmetry returns to be
broken, ${\bar \varphi} \neq 0$,  for a magnetic field above a critical value
given by $eB_c/M^2 \simeq 53.93$.    In {}Fig. \ref{fig6} the effective
potential is plotted for the same fixed temperature of {}Fig. \ref{fig5} and
for values of $B$ below, at and above the critical value $B_c$, above which
the symmetry becomes broken again (as before, to show the spinodal region of
the potential, we have plotted only the real part of it).  It should be
noticed from these results that the phase transition is once  again second
order. The effect of the magnetic field is to enlarge the symmetry  broken
region by making the critical temperature larger, but it does not change the
order of the transition.

\begin{figure}[htb]
  \vspace{0.75cm} \epsfig{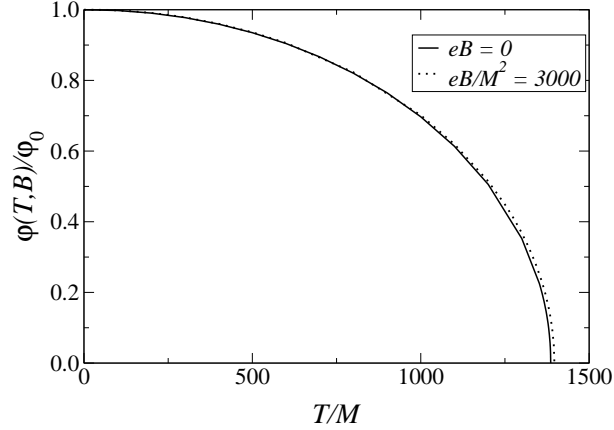}
\caption[]{Temperature dependence of the normalized  VEV ${\bar
    \varphi}(T,B)/\varphi_0$. The critical temperature for $B=0$ is
  $T_c^{B=0}/M \simeq 1385.64$. Including the effects of the magnetic field,
  with $e B/M^2 = 3000$, $T_c$ increase to $T_c/M \simeq 1396.87$.  The other
  parameters used here are:  $\lambda = 0.00375$ and $m/M = 20$.}
\label{fig7}
\end{figure}

{}Finally, we have again used for comparison a case with analogous parameters
as considered in  Ref.~\cite{ayala-ring}, and contrary to the results found in
that reference, we have found again here only a second order phase transition.
{}For the analogous parameters used in \cite{ayala-ring} (see observations
made in previous subsection), $m/M=20$, $\lambda = 0.00375$ and  $eB/M^2 =
30$, the change of the critical temperature in relation to the $B=0$ case
shown in {}Fig. \ref{fig3} is only to produce a slight shift of the critical
temperature to a value $0.08\%$ higher. In {}Fig. \ref{fig7}, we have plotted
the  VEV of the field for  the case of zero magnetic field and for a much
higher value of the magnetic field, $eB/M^2 = 3000$, so to be able to
visualize the difference. Even so, the difference even for such value of
magnetic field is only marginal, changing the critical temperature obtained at
$B=0$, $T_c(B=0)/M \simeq 1385.64$, to $T_c(B)/M \simeq 1396.87$ for the value
of $B$ used.  In all cases we have tested for a nonvanishing magnetic field
and other different parameters of the potential, the transition is again
verified to be second order, with the VEV varying always continuously from its
value at $T=0$ to zero at $T=T_c(B)$. 


\section{Using the Euler-Maclaurin formula for the sum over Landau levels}

When working with quantum field theories in a magnetic field, we need to  deal
with the sum over the Landau levels. At $T=0$ this does not present  a problem
in general, since we can express the expressions in terms of  zeta
functions. We have seen this in the previous section, where all terms at $T=0$
were presented in analytical form. However, at finite temperature this is not
possible in general, therefore, we either need to numerically perform the sums
over the Landau levels, or find suitable approximations  for the
expressions. This job becomes easier in the high magnetic field regime, with
$\sqrt{eB} \gg T$, for which most of the time suffices to use only the very
first terms of the sum, or even just the leading Landau level term.  Higher
order terms are quickly Boltzmann suppressed in this case. But this is not so
in the low magnetic field regime, with  $\sqrt{eB} \lesssim T$, and $\Omega/T
\lesssim 1$,  where most of the time requires to sum up to very large Landau
levels. This has the potential to make  any numerical computation to become
quickly expensively, specially when there are in  addition numerical
integrations involved, like in our case here (note that we have not made use
of high temperature approximations, but worked  with the complete expressions
in terms of the temperature dependent integrals). Simple approximate
analytical  results, which are always desirable to obtain, can also become
difficult to obtain without  a suitable approximation that can be used.  One
such approximation was discussed in Ref.~\cite{ayala-ring}, but it is valid
only for very small magnetic fields. Here we discuss a more natural
alternative  for dealing with the cases of low magnetic fields, based on the
Euler-Maclaurin formula. 

The Euler-Maclaurin (EM) formula provides a connection between integrals and
sums. It can be used to evaluate finite sums and infinite series using
integrals, or vice-versa. In its most general form it can be written
as~\cite{EMidentity}:

\begin{eqnarray}
\sum_{k = a}^{b}f(k) &=& \int_{a}^{b} f(x)\,dx + \frac{1}{2}\left[ f(a)+f(b)
  \right]  + \sum_{i=1}^{n}\frac{b_{2i}}{(2i)!}\left[f^{(2i-1)}(b) -
  f^{(2i-1)}(a)\right] \nonumber \\ &+& \int_{a}^{b}
\frac{B_{2n+1}(\{x\})}{(2n+1)!}  f^{(2n+1)}(x)\,dx\;,
\label{EM}
\end{eqnarray}
where $b_i$ are the Bernoulli numbers, defined by the generating function

\begin{equation}
\frac{x}{\exp(x)-1} = \sum_{n=0}^\infty b_n \frac{x^n}{n!}\;,
\end{equation}
and $B_{n}(x)$ are the Bernoulli polynomials, with generating function

\begin{equation}
\frac{z \exp(zx)}{\exp(z)-1} = \sum_{n=0}^\infty B_n(x) \frac{z^n}{n!}\;.
\end{equation}
The notation $\{x\}$ in $B_{2k+1}(\{x\})$ in Eq. (\ref{EM}) means the
fractional part of $x$ and $f^{(k)}(x)$ means the $k$-th derivative of the
function.  The last term in Eq. (\ref{EM}) is known as the {\it remainder}.

Next, we study the reliability of the use of the EM formula as a suitable
approximation  for the sums over Landau levels and we estimate the errors
involved in doing so.  In the following, we will call the 0th-order in the EM
formula when only the first term  in the RHS of Eq. (\ref{EM}) is kept, the
1st-order when also the second term is kept  and so on so forth. Thus, for
example, the Landau sum part of ${\tilde Y}(T,B,\eta)$ in Eq. (\ref{YTB}), up
to second order in the EM formula, becomes

\begin{eqnarray}
L{\tilde Y}&=&  \sum_{k=0}^{\infty} \int_{-\infty}^{+\infty} \frac{d z}{2 \pi}
\ln \left[ 1- e^{ -E(z,\Omega, T, B, k) } \right]   \simeq
\int_{-\infty}^{+\infty} \frac{d z}{2 \pi}   \left\{ \int_{0}^{\infty}\ln
\left[ 1- e^{-E(z,\Omega, T, B, k)} \right]d k \right.
\nonumber\\ &+&\left. \frac{1}{2} \ln \left[ 1- e^{-E(z,\Omega, T, B, 0)}
  \right]  \right.\nonumber\\ & - &\left. \frac{e B}{12 T^2}\;
\frac{1}{E(z,\Omega, T, B, 0) \left[e^{E(z,\Omega, T, B, 0)}-1\right]}
\right\}\;,
\label{EMY}
\end{eqnarray}
where 

\begin{eqnarray}
E(z,\Omega, T, B, k)  =   \sqrt{z^2 + \frac{\Omega^2}{T^2} + (2k+1)\frac{e
    B}{T^2}}\;.
\end{eqnarray}
Likewise, the Landau sum part of ${\tilde X}(T,B,\eta)$ in Eq. (\ref{XTB}), up
to second order in the EM formula, becomes

\begin{eqnarray}
L{\tilde X}& =&\sum_{k=0}^{\infty} \int_{-\infty}^{+\infty} \frac{d z}{2 \pi}
\frac{1}{E(z,\Omega, T, B, k)} \frac{1}{e^{E(z,\Omega, T, B, k)} -1} \simeq
\int_{-\infty}^{+\infty} \frac{d z}{2 \pi}\left\{\int_0^{\infty}
\frac{1}{E(z,\Omega, T, B, k)}\frac{1}{e^{E(z,\Omega, T, B, k)}-1}\, d k
\right.      \nonumber\\ &+&  \left.\frac{1}{2} \;  \frac{1}{E(z,\Omega, T, B,
  0) \left[e^{E(z,\Omega, T, B, 0)}-1\right]}  \right.\nonumber\\ & +
&\left. \frac{e B}{12T^2}\;  \frac{1}{E^2(z,\Omega, T, B, 0)
  \left[e^{E(z,\Omega, T, B, 0)}-1\right]} \left[\frac{1}{E(z,\Omega, T, B,
    0)} +  \frac{e^{E(z,\Omega, T, B, 0)}}{e^{E(z,\Omega, T, B,
      0)}-1}\right]\right\}\;.
\label{EMX}
\end{eqnarray}

In Tab. \ref{tab1} we assess the reliability of using the EM at each order (we
have  studied it up to the forth-order) for the  sums in Eqs. (\ref{EMY}) and
(\ref{EMX}), for different values for the ratio $eB/T^2$, and we compare the
results with the exact ones. 
  
\begin{table}[!htb]
\begin{tabular}{c|c|c|c|c|c|c|c}
\hline    $eB/T^2$  & $\;$order$\;$ & \multicolumn{3}{c|}{$L{\tilde Y}$} &
\multicolumn{3}{c}{$L{\tilde X}$} \\  \cline{3-8} &               & $\;$ EM
approx.$\;$& $\;$Landau Sum $\;$ & $\;$error ($\%$) $\;$ & $\;$ EM approx.$\;$
&  $\;$ Landau Sum $\;$ & $\;$error ($\%$) \\ \hline & 0 & - 688.7717
&                         & 0.04   & 508.2575              &
& 1.67   \\ & 1 & - 689.0258              &                         &
$1.00\times10^{-4}$   & 515.7328              &                       & 0.22
\\ $0.001$  & 2 & - 689.0271              & - 689.0264              &
$1.00\times10^{-4}$   & 517.0372              & 516.8633              & 0.03
\\ & 3 & - 689.0270              &                         &
$1.00\times10^{-4}$   & 516.7096              &                       & 0.03
\\ & 4 & - 689.0271              &                         &
$1.00\times10^{-4}$   & 517.2016              &                       & 0.07
\\  \hline & 0 & - 68.6567               &                         & 0.35   &
47.7384               &                       & 5.00   \\ & 1 & - 68.8954
&                         &  $5.10\times10^{-3}$   & 49.8997               &
& 0.70   \\ $0.01$  & 2 & - 68.8990               & - 68.8989
&$1.00\times10^{-4}$   & 50.3031               & 50.2488               & 0.11
\\ & 3 & - 68.8988               &                         &
$1.00\times10^{-4}$   & 50.2007               &                       & 0.10
\\ & 4 & - 68.8990               &                         &
$1.00\times10^{-4}$   & 50.3549               &                       & 0.21
\\  \hline & 0 & - 6.6735                &                         & 2.99   &
3.9418                &                       & 14.07  \\ & 1 & - 6.8706
&                         & 0.13   & 4.4852                &
& 2.22   \\ $0.1$    & 2 & - 6.8797                & - 6.8793                &
$5.80\times10^{-3}$   & 4.6037                & 4.5873                & 0.36
\\ & 3 & - 6.8791                &                         &
$2.90\times10^{-3}$   & 4.5725                &                       & 0.30
\\ & 4 & - 6.8796                &                         &
$4.40\times10^{-3}$   & 4.6199                &                       & 0.70
\\  \hline & 0 & - 0.5400                &                         & 18.70  &
0.2195                &                       & 34.89  \\ & 1 & - 0.6497
&                         & 2.18   & 0.3128                &
& 7.21   \\ $1$      & 2 & - 0.6652                &  - 0.6642               &
0.15   & 0.3416                & 0.3371                & 1.34   \\ & 3 & -
0.6637                &                         & 0.07   & 0.3329
&                       & 1.25   \\ & 4 & - 0.6651                &
& 0.14   & 0.3465                &                       & 2.79   \\  \hline &
0 & - 0.0159                &                         & 59.54  & 0.0034
&                       & 66.99  \\ & 1 & - 0.0328                &
& 16.54  & 0.0081                &                       & 21.36  \\ $10$
& 2 & - 0.0407                &  - 0.0393               & 3.56   & 0.0112
& 0.0103                & 8.74   \\ & 3 & - 0.0383                &
& 2.55   & 0.0095                &                       & 7.77   \\ & 4 & -
0.0413                &                         & 5.09   & 0.0124
&                       & 20.39  \\  \hline & 0 & - $6.8468\times10^{-6}$ &
& 88.48  & $5.9362\times10^{-7}$ &                       & 89.52  \\ & 1 & -
$3.6528\times10^{-5}$ &                         & 38.52  &
$3.4235\times10^{-6}$ &                       & 39.54  \\ $100$    & 2 & -
$8.3692\times10^{-5}$ & - $5.9413\times10^{-5}$ & 40.87  &
$8.3706\times10^{-6}$ & $5.6626\times10^{-6}$ & 47.82  \\ & 3 & -
$1.1413\times10^{-5}$ &                         & 80.80  &
$3.6803\times10^{-6}$ &                       & 35.00  \\ & 4 & -
$3.8726\times10^{-4}$ &                         & 551.81 &
$5.6924\times10^{-5}$ &                       & 905.27 \\  \hline
\end{tabular}
\caption{The precision of the EM formula (without the remainder)
  at each order (up to 4th-order),
  compared with the result from the Landau sum. In all cases we have
  considered  $\Omega=0$ in Eqs. (\ref{EMY}) and (\ref{EMX}), when extending them
up to 4th-order.}
\label{tab1}
\end{table}

{}From the results shown in Tab. \ref{tab1}, we see that the EM approximation
produces results that already at first order have good accuracy compared to
the exact values coming from the full Landau sums. It performs particularly
very well for  values of $eB \ll T^2$, reaching convergence within just the
very first few terms in the expansion. However, for values $eB \gtrsim T^2$,
increasing the order in the EM series, it quickly looses accuracy. This can be
traced to the fact that as we  go to a higher order in the derivatives
appearing in the sum on the RHS of Eq. (\ref{EM}), higher powers of $eB/T^2$
are produced, eventually spoiling the approximation (though it oscillates
around the true value). This apparent runaway behavior is only cured by the
introduction of the last term in Eq. (\ref{EM}), the remainder.

It should be noted that Eq. (\ref{EM}) is not an approximation to the sum, but
it is actually an identity. The important term in that context is the
remainder, the last term in Eq. (\ref{EM}). In all our numerical tests, within
the numerical precision for the integrals (for convenience, we have performed
all the numerical integrations  in {\it Mathematica}~\cite{math}), we have
verified that Eq. (\ref{EM}) has agreed with the results from the  sum over
the Landau levels  in the LHS of Eqs. (\ref{EMY}) and (\ref{EMX}), for all
values tested and at all orders (when including the remainder), with a
precision of less than $0.005 \%$. This is quite impressed, recalling that we
can in principle keep only the first few terms in Eqs.~(\ref{EMY}) and
(\ref{EMX}) (the first order approximation) when including the
remainder. {}For numerical  computation this is a tremendous advantage, since
Eq.~(\ref{EM}) can easily be coded, compared to having to perform sums over
very large Landau levels (particularly in the cases of low magnetic fields)
and the multiple integrations that may be required. 

Given all the advantages of using the EM formula, it is quite surprising that
it is not used frequently in physics, despite being well known in mathematics.
In particular, we see that it is well suitable in problems involving external
magnetic fields. To our knowledge, one of the few works that we are aware of
that  have previously made use of the EM formula before were for example
Refs.~\cite{chakra,ho,ravndal}.  In Ref.~\cite{chakra} the author derived the
effective potential of  the abelian-Higgs model, including both thermal
effects and an external magnetic field, in the one-loop approximation.  The EM
formula was used at its zeroth order (by just transforming the sum in an
integral) to obtain analytical results for the effective potential  in the
high magnetic field region. But from the results of Tab.~\ref{tab1}, this is
exactly the region and order in the approximation that lead to the largest
errors.  In \cite{ho} the EM formula was used to obtain an expression for the
effective potential  for vector bosons and to study pair production in an
external magnetic field, while in \cite{ravndal} it was used to obtain
analytical expressions for the internal energies for non-interacting bosons 
confined within a harmonic oscillator potential.  
None of these previous works have assessed the
reliability of the use of the EM formula. 

As an application of the EM formula to our problem, we estimate from it the
dependence of the critical temperature with the magnetic field. Recall from
Eq.~(\ref{varphiTc}) the critical temperature can be obtained by setting
$\Omega=0$ and ${\bar \varphi}=0$ in the equation for the VEV. Thus, from
Eq.~(\ref{varphiTB}), we need to solve the equation

\begin{equation}
6\frac{m^2}{\lambda} -4 {\tilde{X}}(T_c,B)\Bigr|_{\Omega=0}=0\;.
\label{TcBeq}
\end{equation}
As we have seen from the results of Tab. \ref{tab1}, the EM formula  for low
magnetic fields produce reliable results already at the second order, and for
very small magnetic fields, good precision is reached even at first
order. Thus,  for example from the expansion in Eq. (\ref{EMX}), and
performing the $k$ integral in the first term, we obtain

\begin{eqnarray}
L{\tilde X} & \simeq &   -\frac{1}{\pi}\frac{T^2}{eB}\int_{0}^{\infty} d z \,
\ln\left(1-e^{-\sqrt{z^2+a^2}}\right)       
\nonumber\\ &+&  \frac{1}{2 \pi}
\int_{0}^{\infty} d z \,\frac{1}{\sqrt{z^2+a^2}}
\frac{1}{e^{\sqrt{z^2+a^2}}-1} 
\nonumber \\ & + & \frac{e B}{12 \pi T^2}
\int_{0}^{\infty} d z \, \frac{1}{(z^2+a^2) \left(e^{\sqrt{z^2+a^2}}-1\right)}
\left(\frac{1}{\sqrt{z^2+a^2}} +
  \frac{e^{\sqrt{z^2+a^2}}}{e^{\sqrt{z^2+a^2}}-1}\right)\;,  
\label{EMXa}
\end{eqnarray}
where $a^2= eB/T_c^2$. Results for the integrals in Eq. (\ref{EMXa}) can be
found in the  low magnetic field regime, where $a\ll 1$. This regime is
equivalent to the high temperature regime in quantum field theory, from where
a series expansion in $a$ can be found. In fact, the integrals in
Eq. (\ref{EMXa}) can be easily related to the $h_i(a)$ Bose-Einstein integrals
found in that context (see for example the App. A in Ref.~\cite{kapusta}),
from where we obtain the leading order terms in the expansion in powers of
$a$:

\begin{equation}
L{\tilde X}  \simeq  \frac{\pi}{6 a^2} - \frac{5}{24 a} + {\cal O}(a^0)\;.
\label{aprocLXa}
\end{equation}
Going to higher order in the EM series only lead to ${\cal O}(a^n)$ (with
$n\geq 0$) terms,  for $a=\sqrt{eB}/T_c \ll 1$.

Using the result (\ref{aprocLXa}) in Eq. (\ref{XTB}) and then back in
Eq. (\ref{TcBeq}), after some algebra,  we obtain the approximate result for
the critical temperature as a function of the magnetic field,

\begin{equation}
T_c^2(B) \simeq 18 \frac{m^2}{\lambda} + \frac{25 eB}{32 \pi^2}\left( 1+
\sqrt{1 + \frac{1152 \pi^2 m^2}{25 \lambda\, eB}}\right)\;.
\label{TcB}
\end{equation} 

The result (\ref{TcB}) shows the growth of the critical temperature with the
magnetic field, as seen numerically from the results obtained in the previous
section. A comparison with our previous results also shows that
Eq. (\ref{TcB}) provides an excellent fit for  $T_c(B)$,  leading to results
with less than $1\%$ error compared to the full numerical results for $T_c$,
for values of field such that $eB/T_c^2 \lesssim 1$.  {}For example, for the
parameters $\lambda=0.1$ and $eB = 400 m^2$, the approximation (\ref{TcB})
gives $T_c/m \simeq 17.97$, while the full numerical calculation gives $T_c/m
= 18.11$.

\section{Conclusions}

In this paper we have revisited the phase transition problem for the
self-interacting complex scalar field model when thermal effects and an
external magnetic field are present.  We have studied this problem in the
context of the nonperturbative method of the optimized perturbation theory. We
have shown that the OPT method preserves the Goldstone theorem and that
carrying out the approximation to first order is analogous to the ring diagram
resummation method. The OPT carries out the resummation of self-energy
diagrams in a self-consistent way, avoiding overcounting issues, which have
previously plagued the ring diagram method in its very early applications.

By using the OPT method, we have demonstrated that the phase transition in the
model is always second order in the presence of thermal effects and by
including an external  magnetic field, there is no change to the phase
transition order.  The effect of the external magnetic field is to strengthen
the symmetry broken phase, producing a larger VEV for the field  and also a
larger critical temperature. The study we made in this paper considered
only the effects of the magnetic field and temperature on the charged scalar
field (the model is one with a global $U(1)$ symmetry). 
Our study, in a sense, is closer to the case of studying the effects of $B$ and $T$
in the linear sigma model~\cite{andersen}, where no couplings to gauge fields 
(except to the external field) are considered in general.
The inclusion of other field degrees of freedom, or when 
promoting the symmetry from a global
one to a local symmetry,
can of course change both qualitatively and quantitatively 
how the magnetic field affects the system. 
In particular, as concerning the possibility
of producing a first order phase transition, either due to thermal effects, or
by a magnetic field or from both. This is what we expect in the context of
the scalar QED, where due to the screening of the magnetic field in the broken
phase, the phase diagram and transition are very different.
The phase structure can also be very different when adding other interactions 
and having other symmetries. {}For example, in the
context of the electroweak phase transition~\cite{Kajantie,Fiore,Kimmo},
it has been shown that the effect of the magnetic field is to make the first 
order transition stronger. By omitting vacuum energy terms from the 
effective potential, which may have
phenomenological motivations, as in  Ref.~\cite{andersen}, may also lead to
a first order transition, instead of a second order one.

Still comparing our results with those obtained in the abelian Higgs model,
there is a tantalizing question of whether for very strong magnetic fields a
new phase could be formed. This could be for example a phase 
with global vortices condensation, thus restoring the
symmetry again. This would be in analogy to the local, Nielsen-Olesen vortex conndensation
that can happen in type II superconductors~\cite{scalarQED}.
Vortex condensation in that case  is energetically favorable 
to happen for fields beyond a critical value and when the mass of the Higgs 
field becomes larger that the mass of the gauge field. This is an interesting
possibility to investigate in the future.

As an aside, but a complementary part of this work, we have verified the
reliability of the use of the Euler-Maclaurin formula as an approximation for
the sum over Landau energies in a magnetic field. We have verified that it
produces results with errors of less than $0.1\%$ already at the first few
orders in the EM formula and that it leads to  particularly suitable
approximations in the low magnetic field region ($eB/T^2 \ll 1$).  The low
magnetic field region is typically the region where we have to  face the
problem of summing over very larger number of Landau levels, which, in
practice, can be overwhelming in terms of CPU time, when working
numerically. The EM formula in this case can be a very valuable tool, both for
numerical computations, but also for obtaining approximate analytical results,
which would be otherwise very difficult through the use of the Landau sums
directly.  As an application of the EM formula, we have obtained an
approximate  expression for the dependence of the critical temperature with
the magnetic field in our problem. We expect that the EM formula can also be
used in many other applications involving the effects of magnetic fields in
phase transitions, like in condensed matter problems or in the recent interest
in studying the effects of magnetic fields in the QCD phase
transition. {}Furthermore, our results (including the use of the OPT method)
can also be of interest in the study of cosmological phase transitions in
general. Magnetic fields can be easily generated in the early
universe~\cite{magcos}. These fields can then influence subsequent
cosmological phase transitions or also be important in particle physics
phenomena in the early universe.  Works in those contexts are in progress and
we will report on them elsewhere.

\acknowledgments

Work partially supported by CAPES, CNPq and FAPEMIG (Brazilian agencies).
R.L.S.F. would like to thank A. Ayala, E. S. Fraga, A. J. Mizher,
H. C. G. Caldas  and M. B. Pinto for discussions on related matters.

\appendix

\section{Renormalization in the OPT method}
\label{appa}
In this appendix we briefly explain the renormalization  of the effective
potential Eq. (\ref{Veff}). {}First, note that the interpolation procedure in
the OPT method,  Eq. (\ref{interpol}), introduces only new  quadratic terms,
thus, it does not alter the renormalizability of the original theory.
Therefore, the counterterms needed to render the theory finite  have the same
polynomial structure of the original  Lagrangian~\cite{previous1}.

In obtaining the renormalized effective potential, we first note that from the
order $\delta$ term Eq. (\ref{Fc}), we obtain the mass renormalization
counterterm entering in   $\Delta {\cal L}_{\rm ct, \delta}$ in
Eq. (\ref{interpol}),

\begin{equation}
-\frac{1}{2} A_{\delta^1} \phi_i^2 \;,
\label{massct}
\end{equation}
where

\begin{equation}
A_{\delta^1} = \delta \frac{\lambda \Omega^2}{24 \pi^2 \epsilon}\;.
\label{Act}
\end{equation}
{}From Eq. (\ref{massct}), we obtain the two mass counterterms diagrams
contributing to the effective potential at first order in the OPT and shown in
{}Fig. \ref{vacuum}.

By collecting all the divergences from the vacuum loop terms in
Eq. (\ref{Veff}), we have that the divergence from the contribution
(\ref{Fc}) is canceled by the counterterm  Eq. (\ref{massct}). A potential
temperature and magnetic field dependent divergence, proportional to either
$X$ in the two-loop vacuum term Eq. (\ref{Fe}), or ${\tilde X}$ in the case of
a finite external magnetic field, is explicitly canceled against identical
term coming from the mass counterterm diagrams, Eq. (\ref{Fd}). The remaining
divergences in the effective potential are only vacuum ones, independent of
the background field, and they can be all  canceled by introducing a vacuum
counterterm $\Delta V_{\rm ct}$,  added to the effective potential. At the
first  order in the OPT, $\Delta V_{\rm ct}$ is given by 

\begin{equation}
\Delta V_{\rm ct,\delta^1} =  \left(\frac{\Omega^4}{2(4\pi)^2} -
\frac{\delta\eta^2}{16\pi^2}\Omega^2\right)  \frac{1}{\epsilon} +
\frac{\delta\lambda\Omega^4}{3(16\pi^2)^2} \frac{1}{\epsilon^2}\;.
\label{vacuumct}
\end{equation}
Thus, at first order in the OPT, we only require the two counterterms,
Eqs. (\ref{massct}) and (\ref{vacuumct}). Going to second order in the OPT it
will also require a coupling constant renormalization counterterm,
$C_{\delta^2} \lambda \phi_i^4/4!$, from which we can built  vertex
counterterm vacuum diagram contributions to the effective potential~\cite{FGR}
(see also the first two references in \cite{previous1}). At second order in
the OPT we would also obtain $\delta^2$ contributions for Eqs. (\ref{massct})
and (\ref{vacuumct}). Going to higher orders in the OPT only produces
additional  ${\cal O}(\delta^n)$, $n>2$, contributions to the mass, vertex and
vacuum counterterms.


\end{document}